\documentstyle[twoside,fleqn,espcrc2,epsf]{article}

\newcommand{\bea}{\begin{eqnarray}}
\newcommand{\ena}{\end{eqnarray}}

\title{Flavor Singlet Axial Coupling of the Proton -- An Updated
  Analysis \thanks{Poster presented by J. Viehoff.}}
\author{J.~Viehoff$^{\rm a}$ for the SESAM$^{\rm a,b}$ Collaboration
  \\[8pt]
  {\small {\rm $^a$}NIC c/o Forschungszentrum J\"ulich, D-52425
    J\"ulich,
    and DESY, D-22603 Hamburg, Germany.\\
    {\rm $^b$}Physics Department, University of Wuppertal, D-42097
    Wuppertal, Germany.}}

\begin{document}

\begin{abstract}
  We present a combined analysis of SESAM and $T\chi L$ data for the
  flavor singlet axial coupling $G_A^1$ of the proton, which is very
  helpful to stabilize the disconnected signals at small quark masses.
  From connected and disconnected contributions we use the tadpole
  improved renormalization constant $Z_A$ and obtain $G_A^1=0.21(12)$.
\end{abstract}

\maketitle


\section{Introduction}
The flavor singlet axial coupling of the proton is defined as
\begin{equation}
s_{\mu} G_A^1 = \langle P|
\bar{u}\gamma_{\mu}\gamma_5 u + \bar{d}\gamma_{\mu}\gamma_5 d +
\bar{s}\gamma_{\mu}\gamma_5 s| P\rangle,
\label{def}
\end{equation}
where $s_{\mu}$ denotes the components of the proton polarization
vector. In the naive parton model $G_A^1$ is related to the fraction
of the proton spin carried by the quarks. From the measurement of the
first moment of the spin dependent proton structure function, $g_p^1$,
in deep inelastic polarized muon proton scattering the EMC experiment
\cite{EMC_exp} found a small value,
\begin{equation}
G_A^1 = 0.12(17)
\end{equation}
which led to the ``proton spin crisis''.  The result for $G_A^1$
indicates that the contribution to the proton spin from the quarks is
small. New experiments, including proton, neutron and deuteron data
\cite{SMC_exp_new}, have shifted the value up to $G_A^1=0.29(6)$ which
is still far away from the Ellis-Jaffe QCD sum-rule prediction
\cite{ellis_jaffe_sr}:
\begin{equation}
G_A^1\simeq G_A^8=0.579(25)
\end{equation}
Albeit a lot of theoretical and experimental work has been done in the
meantime, a clear understanding of the proton spin is still lacking.
In this context lattice calculations are very illuminating though
hampered by large fluctuations \cite{ga_pap}.

\section{Lattice techniques}

\begin{figure}[htb]
\vspace*{-0.6cm}
\begin{center}
\parbox{7cm}{\epsfxsize=5cm\epsfysize=3cm\epsfbox{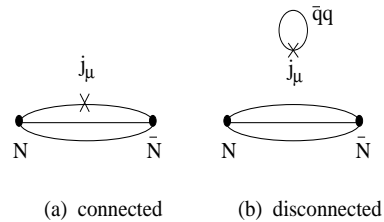}}
\end{center}
\vspace*{-1cm}
\caption[a]{\label{conn_disc}
  \it Connected (a) and disconnected (b) diagrams contribute to the
  flavor singlet axial coupling $G_A^1$ of the proton
  ($j_{\mu}=\bar{q}\gamma_{\mu} \gamma_5 q$). }
\end{figure}

On the lattice both connected and disconnected diagrams for $G_A^1$
(see figure \ref{conn_disc}) can be calculated from ratios of
correlation functions \cite{ga_pap}.

\vspace*{-0.7cm} 

The SESAM collaboration has analyzed 200 $16^3\times 32$ gauge
configurations at each of their four quark masses corresponding to
$m_{\pi}/m_{\rho}=0.833 ...  0.686$. The configurations have been
generated previously in the standard Wilson discretization scheme with
$n_f=2$ mass degenerate quark flavors, at $\beta=5.6$.  In addition
200 $T\chi L$ gauge configurations ($24^3\times 40$ lattices) at the
same coupling $\beta=5.6$ have been analyzed.  The quark mass of the
$T\chi L$ lattices corresponds to the lightest SESAM quark mass.
Details of the simulations can be found in
\cite{txl,sesam_light_spectrum}.

\subsection{Connected contributions}
For the amplitudes from the connected diagrams (see figure
\ref{conn_disc}(a)) in the matrix element, eq.  \ref{def}, we have
applied the global summation method and the insertion technique as
described in \cite{ga_pap}.

From the ratio
\begin{eqnarray}
R_{A_{\mu}}^{SUM}&=& \frac{\sum_{\vec{x}}
\langle P^{\dagger}\sum_{\vec{y},y_0}
\left[\bar{q}\gamma_{\mu}\gamma_5 q\right](\vec{y},y_0)P \rangle }
{\sum_{\vec{x}}\langle P^{\dagger} P \rangle}
\nonumber \\
&-&  \langle\sum_{\vec{y},y_0}
\left[\bar{q}\gamma_{\mu}\gamma_5 q\right](\vec{y},y_0)\rangle\; ,
\end{eqnarray}
where $P$ is an interpolating operator for the proton, the connected
amplitude $C_q=\langle P|\bar{q}\gamma_{\mu} \gamma_5
q|P\rangle_{conn}$ can be obtained from the asymptotic linear slope
\begin{equation}
R_{A_{\mu}}^{SUM}(t) \stackrel{t \rightarrow \infty}{\rightarrow}\;
A + \langle P|\bar{q}\gamma_{\mu}\gamma_5 q|P \rangle\,t \;. 
\label{eq_sum_meth_asympt}
\end{equation}   

\smallskip
\subsection{Disconnected contributions}
For the disconnected amplitudes $D_q=\langle P|\bar{q}\gamma_{\mu}
\gamma_5 q|P\rangle_{disc}$ (see figure \ref{conn_disc}(b)) we have
used the plateau accumulation method (PAM) \cite{ga_pap}: from the
partial summation
\begin{equation}
R_{A_{\mu}}^{PAM}(t,\Delta t_0,\Delta t)
 = \sum_{y_0=\Delta t_0}^{t-\Delta t} R_{A_{\mu}}^{PLA}(t,y_0)\;,
\label{eq_mplateau_meth_def}
\end{equation}  
with
\begin{eqnarray}
R_{A_{\mu}}^{PLA} &=& 
\frac{\sum_{\vec{x}}
\langle P^{\dagger}\sum_{\vec{y}}
\left[\bar{q} \gamma_{\mu}\gamma_5 q\right]
(\vec{y},y_0)
P \rangle }
{\sum_{\vec{x}}\langle P^{\dagger} P \rangle}
\nonumber \\
&-& \langle\sum_{\vec{y}}\left[\bar{q}\gamma_{\mu}\gamma_5 q\right]
(\vec{y},y_0)
 \rangle \; .
\end{eqnarray} 
$D_q$ follows from the asymptotic time dependence ($t\to \infty$) in 
\begin{equation}
R_{A_{\mu}}^{PAM}(t,\Delta t_0,\Delta t) =
B +  D_{q} (t-\Delta t - \Delta t_0) \;.
\label{eq_mplateau_meth_asympt}
\end{equation}

For the calculation of the axial vector quark loops we have used the
spin explicit stochastic estimator technique with complex Z2 noise and
100 stochastic estimates per spin component and gauge field
configuration \cite{ga_pap}.

\section{Raw data}
\begin{figure}[htb]
\vspace*{-3.2cm}
\begin{center}
\parbox{6cm}{\epsfxsize=6cm\epsfbox{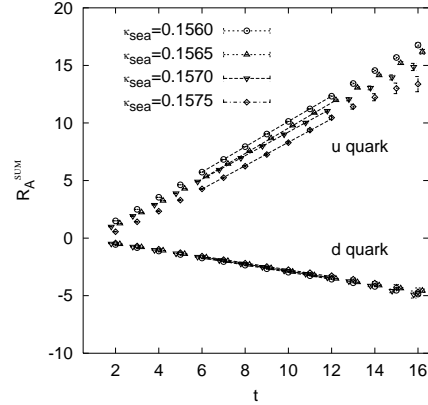}}
\vspace*{-1cm}
\caption[a]{\label{connected_fig} \it $R_{A}^{SUM}(t)$ for 
  the SESAM gauge-configurations (spin-averaged). The linear fits give the
  connected amplitudes $C_u$ and $C_d$.}
\end{center}
\end{figure}

Figure \ref{connected_fig} displays $R_A^{SUM}(t)$ for the connected
amplitudes $C_{u,d}$ on the four SESAM quark masses. The results from
the linear fits are summarized in table \ref{conn_amplitudes}. For the
lightest SESAM quark mass ($\kappa_{sea}=0.1575$) we have also
calculated $C_{u,d}$ on the $T\chi L$ lattice. Note that the volume
effects are less than $7\%$.

\begin{table}[htb]

\begin{center}
\vspace*{-1cm}
\begin{tabular}{|l|l|l|}
\hline
$\kappa$ & $C_u$ & $C_d$\\
\hline
 0.1560 &  1.100(8)  & -0.307(3)  \\
 0.1565 &  1.102(11)  & -0.308(4)  \\
 0.1570 &  1.025(12)  & -0.297(8)  \\
 0.1575(SESAM)  &    1.018(24) & -0.295(10)  \\
 0.1575($T\chi L$) & 1.086(11) & -0.310(6) \\ 
\hline
\end{tabular}
\end{center}
\caption[a]{\label{conn_amplitudes} \it Connected amplitude $C_{u,d}$ from 
the linear fits to $R_{A}^{SUM}(t)$ (SESAM and $T\chi L$).}
\vspace*{-0.7cm}
\end{table}
 
\begin{figure}[htb]
\begin{center}
\vskip -3.0cm
\noindent\parbox{5cm}{
\parbox{5cm}{\epsfxsize=5cm\epsfbox{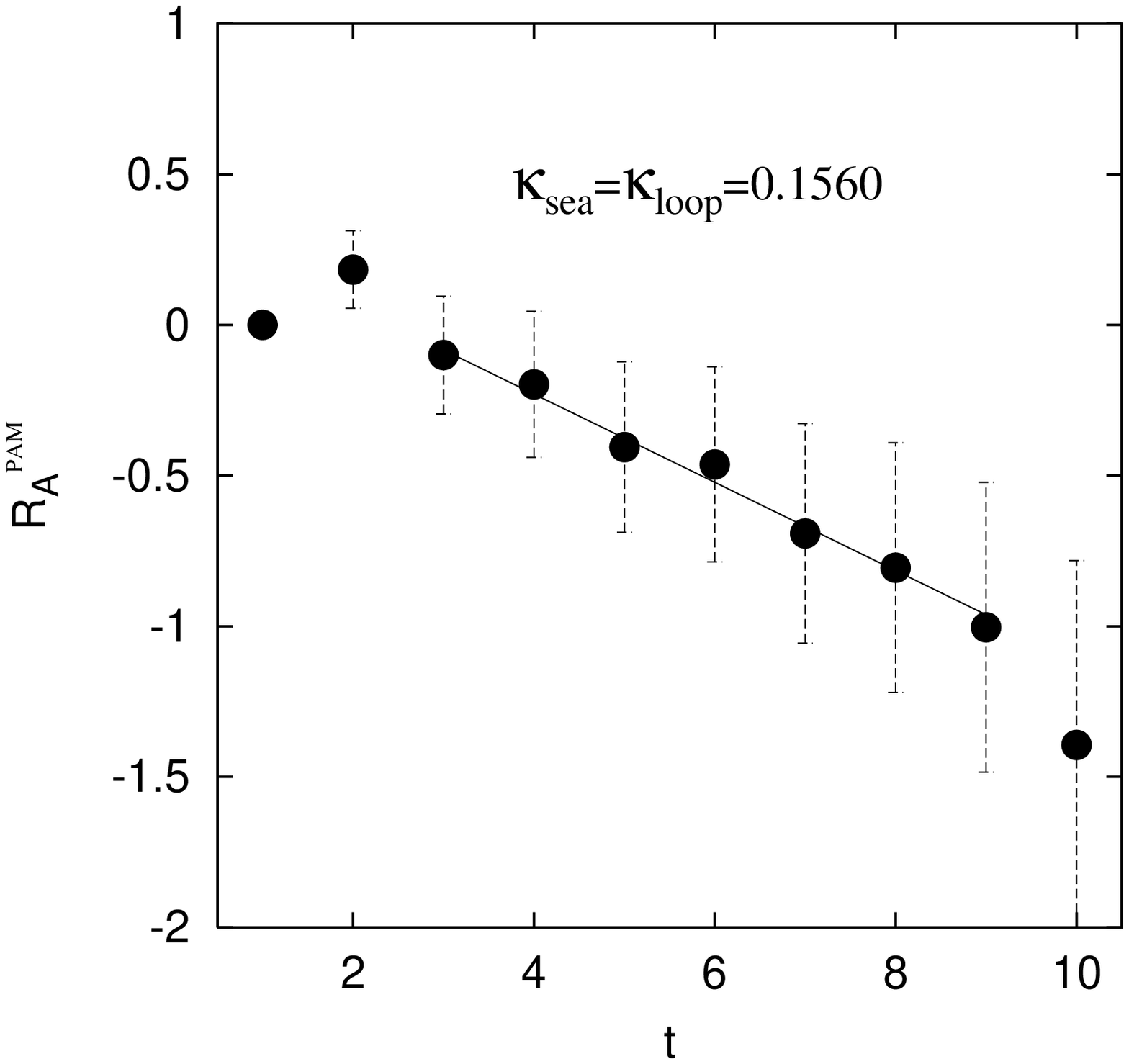}}
\\ \linebreak
\vskip -3.5cm
\parbox{5cm}{\epsfxsize=5cm\epsfbox{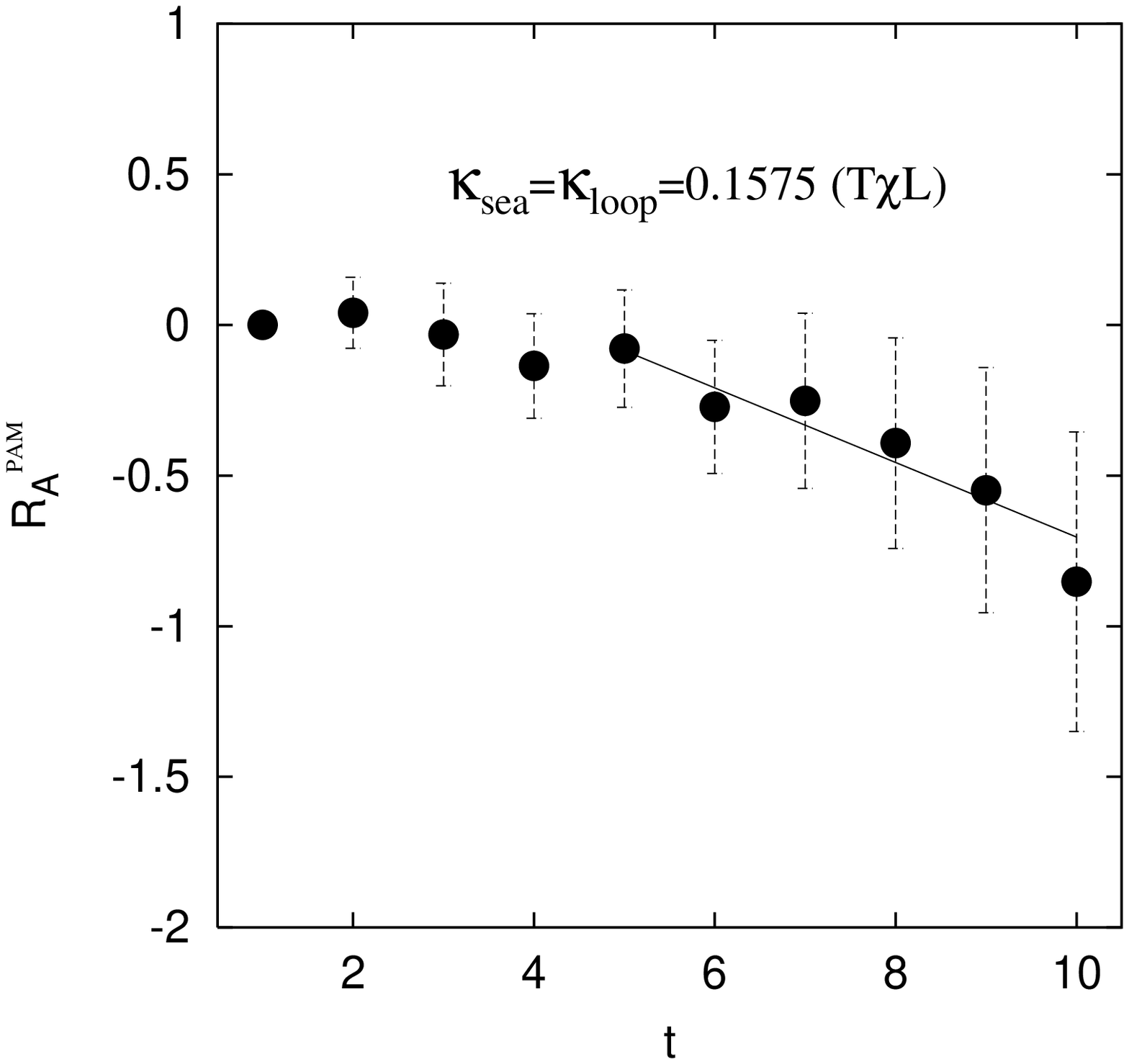}}
}
\vspace*{-1cm}
\caption[a]{\label{disconnected_fig} \it $R_{A}^{PAM}(t)$ for $\kappa_{sea}=0.156$
  (SESAM) and $\kappa_{sea}=0.1575$ ($T\chi L$) (spin-averaged). The
  linear fits give the disconnected amplitude $D_q$.}
\end{center}
\vspace*{-1cm}
\end{figure}

In Figure \ref{disconnected_fig} we plotted $R_A^{PAM}(t)$ ($\Delta
t=\Delta t_0=1$) for the disconnected amplitudes with
$\kappa_{sea}=0.156$ (SESAM) and $\kappa_{sea}=0.1575$ ($T\chi L$).
Albeit the statistical errors for $D_q$ are large we observe comparable 
quality of signals from both plots.

\section{Results}
\begin{figure}[htb]
\vspace*{-3.3cm}
\begin{center}
\noindent
\parbox{7cm}{\epsfxsize=7cm\epsfbox{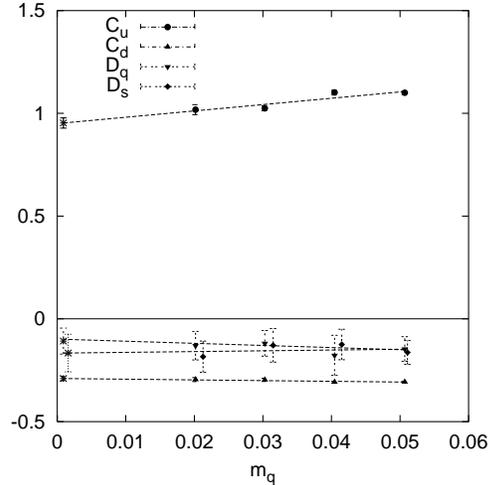}}
\vspace*{-0.9cm}
\caption[a]{\label{chiral_extrap} \it Linear chiral extrapolations of
$C_{u,d}$ and $D_{q,s}$ to the light quark mass $m_{light}$. }
\end{center}
\vspace*{-0.8cm}
\end{figure}

The chiral extrapolations of $C_{u,d}$ and $D_{q,s}$ to the light
quark mass $m_{light}$ are shown in figure \ref{chiral_extrap}.  Since
the finite volume effects are small we have taken the $T\chi L$ data
at $\kappa_{sea}=0.1575$ for the extrapolation in $D_q$. Within the
errors the flavor symmetry $D_q=D_s$ in the disconnected amplitudes is
preserved and we use only the connected amplitudes for $G_A^8$ and
$G_A^3$.

With the Wilson discretization both the flavor singlet and non-singlet
current need renormalization. We have used the first order (tadpole
improved) result from lattice perturbation theory for $Z_A^S$ and
$Z_A^{NS} \cite{ga_pap}$. The renormalized coupling constants of the
proton and the contributions $\Delta q$ to the proton spin are listed
in table \ref{renorm_coupling}.

\begin{table}[htb]
\vspace*{-0.4cm}
\begin{center}
\begin{tabular}{|c|c|c|}
\hline
$\Delta u$ & $\Delta d$ & $ \Delta s$ \\
\hline
 0.62(7)  & -0.28(6)  & -0.12(7)    \\
\hline\hline
 $G_A^1$ & $G_A^3$  & $G_A^8$\\
\hline
 0.21(12) & 0.907(20) & 0.484(18)\\
\hline
\end{tabular}
\end{center}
\caption[a]{\label{renorm_coupling} \it Renormalized coupling constants 
of the proton and the 
   contributions $\Delta q$ to the proton spin.}
\vspace*{-0.8cm}
\end{table}
 

\section{Conclusion}
We have calculated connected and disconnected contributions to the
flavor singlet axial vector coupling of the proton in a full QCD
$n_f=2$ lattice simulation with Wilson fermions. We have checked the
volume effects for the connected amplitudes when we increase the
lattice size at the same coupling ($\beta=5.6$) and quark mass.  In
the analysis of the disconnected contributions we have replaced the
SESAM results for the lightest quark mass by the $T\chi L$ data and
obtained a better signal for $D_q$. Nevertheless, the errors are
dominated by the large statistical fluctuations in the disconnected
amplitudes. We find $G_A^1=0.21(12)$ to be consistent with the result
from experiment and with previous quenched estimates
\cite{japan_ga1,liu_ga1}. For the triplet coupling we get
$G_A^3=0.907(20)$ which is 30\% smaller than the experimental value,
$G_A^3=1.2670(35)$ \cite{SMC_exp_new}. To fix the systematic errors
in the lattice calculation we would need a scaling analysis as well as
a non-perturbative determination of $Z_A^{NS}$ and $Z_A^{S}$. q


\end{document}